\documentclass{aa}
\usepackage{natbib}
\bibpunct{(}{)}{;}{a}{}{,} 
\usepackage{graphicx}
\usepackage{hyperref}
\usepackage{array}
\usepackage{amsfonts}
\usepackage{amssymb}

\begin{document}

  \title{Simulations of micrometeoroid interactions with the Earth atmosphere}

   \author{G. Briani \inst{1,2}
		   \and E. Pace \inst{1}
		   \and S. N. Shore \inst{3,4}
		   \and G. Pupillo \inst{5}
		   \and A. Passaro \inst{6}
		   \and S. Aiello \inst{1} }

   \offprints{G. Briani, email: giacomo.briani@csnsm.in2p3.fr}

  \institute{Dipartimento di Fisica e Astronomia, Universit\`{a} di Firenze, Largo Fermi 2, 50125 Firenze - Italy
	\and present address: Centre de Spectrom\'{e}trie Nucl\'{e}aire et de Spectrom\'{e}trie de Masse, UMR 8609 Universit\'{e} Paris Sud/CNRS-IN2P3, b\^{a}timent 104, 91405 Orsay Campus - France
	\and Dipartimento di Fisica, Universit\`{a} di Pisa, Largo Pontecorvo 3, 50127 Pisa - Italy
	\and Istituto Nazionale di Fisica Nucleare (INFN), sezione di Pisa, Largo Pontecorvo 3, 50127 Pisa - Italy
	\and INAF - Istituto di Radioastronomia, Via Gobetti, 101, 40129 Bologna - Italy
	\and Alta S. p. A., Via A. Gherardesca 5, 56121, Ospedaletto, Pisa - Italy}

\date{Received /
		Accepted }

\abstract{}
{Micrometeoroids (cosmic dust with size between a few $\mu$m and $\sim$1 mm) dominate the annual extraterrestrial mass flux to the Earth.
We investigate the range of physical processes occurring when micrometeoroids traverse the atmosphere.  We compute the time (and altitude) dependent mass loss, energy balance, and dynamics to identify which processes determine their survival for a range of entry conditions.}
{We develop a general numerical model for the micrometeoroid-atmosphere interaction. The equations of motion, energy, and mass balance are simultaneously solved for different entry conditions (e.g. initial radii, incident speeds and angles).
Several different physical processes are taken into account in the equation of energy and in the mass balance, in order to understand their relative roles and evolution during the micrometeoroid-atmosphere interaction.
In particular, to analyze the micrometeoroid thermal history we include in the energy balance: collisions with atmospheric particles, micrometeoroid radiation emission, evaporation, melting, sputtering and kinetic energy of the ablated mass.}
{Low entry velocities and grazing incidence angles favor micrometeoroid survival. 
Among those that survive, our model distinguishes (1) micrometeoroids who reach the melting temperature and for which melting is the most effective mass loss mechanism, and (2) micrometeoroids for which ablation due to evaporation causes most of the the mass loss.
Melting is the most effective cooling mechanism.
Sputtering-induced mass loss is negligible.}
{}


\maketitle

\section[Introduction]{Introduction}
\label{sec:intro}

Micrometeoroids (hereafter referred to as $\mu$METs) with dimensions between $\sim$25 $\mu$m and $\sim$1 mm represent (by mass) the majority population of Solar system minor bodies at a distance of 1 AU from the Sun \citep{grunICARUS85} and dominate the annual mass flux of extraterrestrial matter entering the Earth's atmosphere, i.e. $4 \pm 2\times10^7$ kg/yr, \citep{loveSCIENCE93}.
$\mu$METs can survive the interaction with the atmosphere and reach the Earth surface, becoming micrometeorites.
A large number of micrometeorites, $\geq$ $10^5$, have been recovered in Greenland \citep{mauretteNATURE1987} and Antarctica \citep{mauretteNATURE91, engrandMAPS98, dupratASR07, cordierGCA2011, vanginneken2012MAPS} over the past three decades.  
Many of these recovered $\mu$METs are, however, less altered than expected: unmelted micrometeorites account for between 10 and 30\% for particles $>$100 $\mu$m, $\sim$50\% for sizes between 50 and 100 $\mu$m and up to 78\% in the size range 25-50 $\mu$m \citep{gengeMAPS08}.
In the CONCORDIA collection, which comprises 
micrometeorites collected in 2000, 2002 and 2006 at Dome C, Antarctica, unmelted micrometeorites represent between 33 and 36\% of all the particles 
\citep{dobricaMAPS09,dobricaMETSOC2010}.

When interacting with the Earth's atmosphere, $\mu$METs are subjected to effects of different physical processes.  These include collisions with atmospheric atoms and molecules, deceleration, temperature variations, ablation of $\mu$MET mass and sputtering, and excitation and ionization of atmospheric atoms and molecules.
Several models have been proposed to analyze different aspects of the $\mu$MET - atmosphere interaction. In many cases, models for the $\mu$MET-atmosphere interaction focus on specific processes or are aim to explain particular meteor observations.  
\citet{flynnLPSC1989} modeled the $\mu$MET - atmosphere interaction using the U.S. Standard Atmosphere values for the atmospheric density instead of the exponential approximation proposed by \citet{whipplePNAS1950} and \citet{fraundorfGRL1980}, but assuming that the $\mu$MET mass remains constant during the interaction.
\citet{loveICARUS91} analyzed the heating experienced by $\mu$METs.  Their model includes vaporization and ablation of the $\mu$MET material but not sputtering, and they did not analyze how the different processes evolve during the $\mu$MET atmospheric flight.
\citet{campbell-brownAA04} tuned their model to explain light curves of faint meteors ($\mu$METs in the size range 10 $\mu$m - 2 mm).  In particular, they modeled the $\mu$MET fragmentation to explain sudden increases observed in the light curves and to characterize the size of the grains that constitute the $\mu$METs.
\citet{rogersPSS05} compared the mass loss due to sputtering to that due to ablation.  They assumed,however, that all the energy transferred to a $\mu$MET by the collisions with atmospheric particles can be used by sputtering, rather than only a fraction of it, as should be the case because other physical properties (temperature increment, evaporation, etc.) depend on the collision energy.
\citet{szaszEMP2008} added the treatment of sputtering proposed by \citet{tielensApJ94} to the model of \citet{loveICARUS91}  to model $\mu$MET head echoes recorded by the UHF European Incoherent Scatter facility.
\citet{vondrakACP2008} proposed a detailed model of the differential ablation of the $\mu$MET elements.  They included thermal ablation (i.e. vaporization and melting) and sputtering and focused their analysis on the height at which the different $\mu$MET chemical  constituents (e.g. Na, Mg, Si, Ca, Fe) are released in the atmosphere.

We approach the problem of the $\mu$MET - atmosphere interaction in a holistic way, considering those physical processes that can be the most important and highlighting their relative contributions with different atmospheric entry conditions and their evolution during the $\mu$MET atmospheric flight.
This work is, however, not an exhaustive analysis of the $\mu$MET-atmosphere interaction.  Our goals are (a) to highlight the necessity of a general study (b) to propose a model that includes different physical processes and to analyze their respective contributions and (c) to delineate the most critical aspects of the problem.

\section[The model: properties of micrometeoroids and present-day atmosphere]{The model: properties of micrometeoroids and\\present-day atmosphere}
\label{sec:model_MM}

We treat $\mu$METs as homogeneous, constant density spheres, with radius $r_\mathrm{\mu MET}$. Their initial radii are assumed to be between 25 $\mu$m and 500 $\mu$m.
The upper limit of 500 $\mu$m is required by the assumption of isothermality: we assume that a $\mu$MET is sufficiently small to achieve a uniform internal temperature $T_\mathrm{\mu MET}$ through thermal conduction.
Actually, thermal gradients develop within $\mu$METs due to endothermic reactions [e.g. phase transitions, see \citet{flynnLPSC95}] and are observed in recovered micrometeorites \citep{gengeGCA06} as well as in experimental simulations \citep{toppaniMAPS01}.
But a treatment of $\mu$MET internal temperature distribution is beyond the scope of the present work (a qualitative discussion is presented in section \ref{sec:discussion}).
Since we also assume constant emissivity, the $\mu$MET size must be larger than the wavelength of the emitted and absorbed radiation.  A $\mu$MET generally absorbs radiation in the visible and ultraviolet (VIS-UV) in the spctral range 200 - 800 nm and emits infrared (IR) radiation (at temperatures of $\sim$1000 K) with wavelength $\approx$ 3 $\mu$m \citep{coulsonMNRAS03}, so the simulations terminate when $r_\mathrm{\mu MET} < 5 \mu$m. 
Micrometeoritic structure and compositions are particularly difficult issues.
To treat the thermal processes occurring during their atmospheric passage \citep{greshakeMAPS98,gengeMAPS08}, we assume the $\mu$METs have the composition of carbonaceous chondritic micrometeorites (which represent $\sim 84\%$ of the recovered micrometeorites, see e.g. \citealt{gengeLPSC2006} and \citealt{levisonNATURE2009}). In particular, we assume that $\mu$METs are predominantly silicates. We choose the values of their thermodynamic parameters accordingly, as described below.
As specified above, the $\mu$MET density $\rho_\mathrm{\mu MET}$ is assumed to be constant during the atmospheric flight.
We used two different values for $\rho_\mathrm{\mu MET}$, 3 and 1 g/cm$^3$, to span the range of density measured for recovered micrometeorites \citep{gengeMAPS08}.
Similar values are adopted in \citet{campbell-brownAA04} and \citet{rogersPSS05}.

The present day Earth atmosphere is represented by the altitude profiles of density and the concentration of neutral components from the MSISE-90 model \citep{hedinJGR91}.  We include six atmospheric species: N$_2$, O, O$_2$, Ar, He, H.  We neglect the effects of charge accumulation in our models and therefore do not include atmospheric ions.  Such processes are left for a future development.

The atmospheric properties determine also how a $\mu$MET moves in the atmosphere.  Because the mean free path of an atmospheric particle, at heights between 60 and 200 km (those of interest in this work), is always longer than the $\mu$MET radius, we apply the free molecular flow regime.
In other words, $\mu$METs always interact with single atmospheric atoms and molecules, and there is not a gas layer that protects them from collisions \citep{sorasioPSS01, campbell-brownAA04}.

\section[The model equations]{The model equations}
\label{sec:model_equations}

The two dimensional micrometeroid trajectory is described by the time dependent $h$ (height from the Earth surface) and $z$ (angle with respect to the zenith direction). The equations describing their evolution are:
\begin{equation}
\frac{dh}{dt}=-v_\mathrm{\mu MET}\cos z
\label{eq_height}
\end{equation}
and
\begin{equation}
\frac{dz}{dt}=-\frac{\sin z}{v_\mathrm{\mu MET}}\left(g(h)-\frac{v_\mathrm{\mu MET}^2}{R_{\oplus}+h}\right)
\label{eq_z_angle}
\end{equation}
where $v_\mathrm{\mu MET}$ is the $\mu$MET velocity, $g(h)$ is the gravitational acceleration at height $h$ and $R_{\oplus}$ is the radius of the Earth.
The initial value, $h = 400$ km, applies to the present-day terrestrial atmosphere: the ambient density above this height is negligible. For the entry angle, different values were simulated (0$^{\circ}$, 45$^{\circ}$ and 70$^{\circ}$); for angles greater than 70$^{\circ}$, $\mu$METs tend to skip back to space.  The initial value ($v_\mathrm{in}$) of the $\mu$MET velocity $v_\mathrm{\mu MET}$ was left as a free parameter, the observed range is from  11.2 km/s to 72.8 km/s \citep{ceplechaSSR98}.

Collisions with atmospheric atoms and molecules determine the $\mu$MET velocity $v_\mathrm{\mu MET}$. We model this neglecting the thermal velocity of atmospheric particles ($\sim$370 m/s for $T_\mathrm{atm}$ = 160 K at $h$ = 100 km) relative to $v_\mathrm{\mu MET}$ using the equation of motion:
\begin{equation}
\frac{dv_\mathrm{\mu MET}}{dt}=-\frac{\Gamma}{M_\mathrm{\mu MET}}S\rho_\mathrm{atm}(h)v_\mathrm{\mu MET}^2 + g(h)\cos z
\label{eq_motion}
\end{equation}
where $S=\pi r_\mathrm{\mu MET}^2$ is the $\mu$MET geometric cross section, $M_\mathrm{\mu MET}$ is its mass, and $\rho_\mathrm{atm}(h)$ is the altitude dependent atmospheric density.  The drag coefficient $\Gamma$ describes the momentum transfer efficiency in a single collision to a $\mu$MET; its value is $0 \le \Gamma \le 2$ ($\Gamma$=2 for elastic collisions) and we assume $\Gamma$ = 1 \citep{campbell-brownAA04, rogersPSS05}.

The core of our model is the description of how $\mu$METs acquire energy and of the changes thus produced.  $\mu$METs acquire energy by collisions with atmospheric gas.  Since $v_\mathrm{th} \ll v_\mathrm{\mu MET}$, the heating rate is given by:
\begin{equation}
\frac{dE_\mathrm{coll}}{dt}=\frac{\Lambda}{2}S\rho_\mathrm{atm}(h)v_\mathrm{\mu MET}^3
\label{eq_incoming_energy}
\end{equation}
where $\Lambda$, heat transfer coefficient, parameterizes the fraction of the kinetic energy of an incoming atmospheric particle that is transferred to the target $\mu$MET. 
Its value depends on the $\mu$MET temperature and on the $\mu$MET Mach number ${\cal M}_{\mu MET} \equiv v_\mathrm{\mu MET}/v_\mathrm{th}$ \citep{hoodICARUS91,meloshLPSC08}. 
However, in the limit where ${\cal M}_{\mu MET} \gg 1$, $\Lambda \approx 1$ (see section \ref{sec:discussion} for a discussion) and we therefore assume $\Lambda$ = 1, as do other recent studies (e.g. \citealt{campbell-brownAA04, rogersPSS05}).  Two other possible heating sources are solar irradiation \citep{mosesICARUS92,mcauliffeICARUS06}, and atmosphere radiation. 
We did not include these sources in our energy balance because they are negligible for bodies as small as $\mu$METs.  We estimated the contributions of solar and atmosphere radiation for $r_\mathrm{in}$ = 500 $\mu$m, $v_\mathrm{in}$ = 11.2 km/s, $z_\mathrm{in}$ = 45$^{\circ}$, $\rho_\mathrm{\mu MET}$ = 3 g/cm$^3$.  The energy is 2 orders of magnitude larger than that gained from collisions at $h$ = 262 km and 1 order of magnitude larger at $h$ = 170 km.  However, this energy acquired by radiation does not provoke either a reduction in the size of the $\mu$MET size or a temperature increase.  Energy acquired by radiation becomes negligible, as in previous models \citep{mosesICARUS92}, when denser atmospheric layers are reached where the collisions dominate.  In our example ($r_\mathrm{in}$ = 500 $\mu$m, $v_\mathrm{in}$ = 11.2 km/s, $z_\mathrm{in}$ = 45$^{\circ}$, $\rho_\mathrm{\mu MET}$ = 3 g/cm$^3$) collisions start to dominate at $h$ = 128 km and the maximum energy acquired by collisions overcomes that from radiation by 3 orders of magnitude.  Thus, the $\mu$MET's temperature is regulated by the collisions: their peak temperatures and final sizes do not change when radiation effects are included.

The rise in temperature due to collisional heating drives evaporation and melting of the $\mu$MET, which reduce its mass and dimensions. Here the word ``evaporation'' refers to sublimation and that mass is assumed to be lost immediately.  We assume that the evaporation begins as soon as $\mu$METs start to acquire energy from collisions \citep{mosesICARUS92,campbell-brownAA04, rogersPSS05} with the temperature dependent mass loss rate being modeled using the Knudsen-Langmuir formula \citep{Bronshten83}:
\begin{equation}
\left(\frac{dM_\mathrm{\mu MET}}{dt}\right)_\mathrm{evap} = S
\sqrt{\frac{\mu}{2\pi k_\mathrm{B}T_\mathrm{\mu MET}}}
\times 10^{(A-B/T_\mathrm{\mu MET})}
\label{eq_mass_loss_evap}
\end{equation}
where $k_\mathrm{B}$ is the Boltzmann constant.
Here the relation $\log_\mathrm{10}(p_\mathrm{sat})=A-B/T_\mathrm{\mu MET}$ (solution of the Clausius-Clapeyron equation) is used to evaluate the $\mu$MET saturated vapor pressure.  We assume $A$ = 10.6 and $B$ = 13500 K (\citep{loveICARUS91}).  
We adopt $\mu$ = 45 amu for the mean atomic mass of the $\mu$MET constituents that is appropriate for a chondritic composition \citep{loveICARUS91}. 

Melting also drives mass loss.  At high temperatures, between $\sim$1500 and $\sim$2000 K, portions of the body undergo pass from a solid to a liquid and we assume that this melted mass is also completely lost.  
We note, however, that the very complex composition of chondritic $\mu$METs means that they have a range of melting points, so different parts within a $\mu$MET can begin melting while others remain solid.  
We model melting using a term proportional to the heating, as suggested by \citet{campbell-brownAA04}.  To enforce the condition that the rate of mass loss due to melting is different from zero only if $T_\mathrm{\mu MET} \approx T_\mathrm{melt}$ = $\mu$MET melting temperature, we employ a  parametric coefficient $P_\mathrm{spall}$, defined as
$$
P_\mathrm{spall}=0.5 + \frac{\arctan(T_\mathrm{\mu MET}-T_\mathrm{fus})}{\pi} \;\;\; \mathrm{for} \;\;\; T_\mathrm{\mu MET} < T_\mathrm{melt}
$$
$$
P_\mathrm{spall}=0.5 + \frac{\arctan[4(T_\mathrm{\mu MET}-T_\mathrm{fus})]}{4\pi} \;\;\; \mathrm{for} \;\;\; T_\mathrm{\mu MET} > T_\mathrm{melt}.
$$
This coefficient was introduced by \citet{campbell-brownAA04}, but in their work its value rapidly reaches 1 for $T_\mathrm{\mu MET} > T_\mathrm{melt}$, (i.e. all the heat is used to melt the $\mu$MET mass).    Since melting is only one of the processes included in the energy balance (\ref{eq_energy_balance}), we adopt a form of $P_\mathrm{spall}$ that limits its value to 0.625 (the two different expressions make $P_\mathrm{spall}$ a continuous function of $T_\mathrm{\mu MET}$, to avoid problems in the numerical integration).
The assumed value for the $\mu$MET melting temperature is $T_\mathrm{melt}$= 1623 K. This is similar to values adopted in previous studies  \citep{bonnyLPSC90, loveICARUS91, flynnACM92, scarsiNUOVOCIM04} and is the melting point of silicates that are the dominant phases in $\mu$METs.  The rate of mass loss caused by melting is thus given by
\begin{equation}
\left(\frac{dM_\mathrm{\mu MET}}{dt}\right)_\mathrm{melt}=\frac{P_\mathrm{spall}}{H_\mathrm{melt}}\left(\frac{dE_\mathrm{coll}}{dt}\right)
\label{eq_mass_loss_fus}
\end{equation}
where $H_\mathrm{melt} = 2.65\times10^5$ J/kg is the latent heat of melting for stony meteoroids \citep{loveICARUS91}.

Sputtering is the ejection of atoms or molecules (targets) by collisions with incident particles (projectiles) of sufficient kinetic energy.  This mechanism has been proposed to explain high altitude meteors \citep{brosch01} and previous studies suggest that it can cause significant mass loss \citep{rogersPSS05}.  We require the sputtering yield $Y$, i.e. the number of target atoms or molecules removed by each projectile impact.  To calculate $Y$ we chose a semi-empirical formula from \citet{draineApJ79}, again using $\mu$ = 45 amu as mass of the target particles and U$_0$ = 5.7 eV as their binding energy (this value is that indicated for a silicate compound in \citealt{draineApJ79}).  Using $s$ for the total number of atmospheric species considered, the rate of mass loss by sputtering is:
\begin{equation}
\left(\frac{dM_\mathrm{\mu MET}}{dt}\right)_\mathrm{sputt}=\mu S v_\mathrm{\mu MET}\sum_\mathrm{i=1}^\mathrm{s} n_\mathrm{i} Y_\mathrm{i}
\label{eq_mass_loss_sputt}
\end{equation}
where $n_\mathrm{i}$ is the number density of the i$^\mathrm{th}$ atmospheric species.

In summary, evaporation, melting, and sputtering are considered in the energy balance.   As emphasized by \citet{loveICARUS91}, it is essential to include all three to properly evaluate the $\mu$MET temperature.  An important feature of our model is that the contributions of melting and evaporation are evaluated individually.  The problem is to estimate the fraction $\eta$ of the incoming energy that is used for sputtering and not for the other processes. \citet{popovaASR07} proposed that particles removed by sputtering have a kinetic energy varying from 0\% of the energy supplied by collisions (when the $\mu$MET entry velocity is $<$ 20 km/s) to 20\% (when $\mu$METs enter the atmosphere at 70 km/s), and that the rest of the energy contributes to the rise of $\mu$MET temperature that causes the emission of radiation and ablation.  Based on this, and considering that a part of the incoming energy must be used to remove these particles, we assume $\eta$  = 0.15 for every value of the $\mu$MET entry speed.

Beyond processes responsible for the mass loss, the rise of $\mu$MET temperature causes intense emission of radiation.  This cooling process is fundamental for the $\mu$MET thermal balance, unlike larger meteoroids. As described in section \ref{sec:model_MM}, we assume $\mu$METs have a constant emissivity, $\epsilon = 0.9$, as suggested in \citet{campbell-brownAA04}.

Finally we add a term for the kinetic energy of the ablated mass. As proposed in \citet{campbell-brownAA04}, we assume that the ablated $\mu$MET mass leaves with a speed determined by the $\mu$MET temperature $T_\mathrm{\mu MET}$ .
The energy balance is thus given by:

$$
\frac{\Lambda}{2} S\rho_\mathrm{atm}(h)v_\mathrm{\mu MET}^3 = cM_\mathrm{\mu MET}
\frac{dT_\mathrm{\mu MET}}{dt} +4\pi r_\mathrm{\mu MET}^2 \epsilon \sigma T_\mathrm{\mu MET}^4+
$$
$$
+H_\mathrm{evap}\left(\frac{dM_\mathrm{\mu MET}}{dt}\right)_\mathrm{evap}
+H_\mathrm{melt}\left(\frac{dM_\mathrm{\mu MET}}{dt}\right)_\mathrm{melt}+
$$
\begin{center}
\begin{equation}
+\frac{\Lambda}{2} Sv_\mathrm{\mu MET}^3\sum_\mathrm{i=1}^\mathrm{s}(\eta_\mathrm{i}n_\mathrm{i}m_\mathrm{i})
+\frac{1}{2}\left(\frac{dM_\mathrm{\mu MET}}{dt}\right)_\mathrm{abl}\frac{3k_\mathrm{B}T_\mathrm{\mu MET}}{\mu}
\label{eq_energy_balance}
\end{equation}
\end{center}
where
$\sigma$ is the Stefan-Boltzmann constant, $\eta_i$, $n_i$ and $m_i$ are the sputtering coefficient, number density and mass, respectively, of the atmospheric components.  The expression for the energy used in sputtering takes into account that, depending on the collision velocity, sputtering is caused by only some of the atmospheric species (i.e. if $Y$ is zero for the i$^\mathrm{th}$ species, then $\eta_i = 0$).  
For the thermodynamic properties of $\mu$METs we adopt values used in \citet{Bronshten83} and \citet{loveICARUS91} for stony meteoroids: c = $10^3$ J/(kg K) is the $\mu$MET specific heat, $H_\mathrm{evap} = 6.05\times10^6$ J/kg is the latent heat of vaporization.  
The variation of each term in equation \ref{eq_energy_balance} is followed throughout the numerical integrations so it is possible to see which are most responsible for determining the $\mu$MET fate and valuate how their roles change with different entry conditions and time during flight.

\section[Numerical integration and validation of our code]{Numerical integration and validation of our code}
\label{sec:numerical_integr}

Numerical integrations were performed using the RK4 method with an adaptive step size. For each quantity  (i.e., $\mu$MET height, angle with respect to the vertical, speed, radius, mass and temperature) we required a minimum relative variation of $5 \times 10^{-5}$, and the maximum relative variation allowed was $10^{-3}$.  
If the relative variations were too small, the integration step size is doubled with an imposed upper limit of 1 s; if the variations are too large then the calculated values are rejected and the integration step size halved with a lower limit of $10^{-10}$ s.  
We verified that the code reproduced the results of previous studies.  Several temperature vs time profiles are shown in  \citet{loveICARUS91}. 
They use different initial temperatures and atmospheric properties than we do but we choose the difference between $\mu$MET peak temperatures as parameter for a comparison (i.e. the difference between the peak temperatures of $\mu$METs that have equal entry speed, angle and density, but with different initial size).  
Introducing the set of equations of \citet{loveICARUS91} in our code, we obtain peak temperature differences that differ from the original ones for less than 9\%.  A second comparison was made with \citet{rogersPSS05}, who focused on the sputtering-induced mass loss.  
We computed the fractional mass lost by sputtering for $\mu$METs with $\rho_\mathrm{\mu MET} = 3$ g/cm$^3$ and for $\mu$METs with $\rho_\mathrm{\mu MET} = 1$ g/cm$^3$, considering value of entry speed ranging between 20 and 70 km/s  using the same set of equations of \citet{rogersPSS05}. 
The maximum difference between our results and those of \citet{rogersPSS05} is 6\%.

\section[Results]{Results}
\label{sec:results}

Our simulations highlight how decisive entry conditions are
\footnote{We presented preliminary results of our model in a previous work \citep{brianiMSAItS07}.
In the present paper we correct a few errors in the model equations.  We changed the expression for $P_\mathrm{spall}$ so that its maximum value is 0.65; we changed the equation for the
rate of mass loss due to sputtering (\ref{eq_mass_loss_sputt}) so that we now include sputtering caused only by those atmospheric molecules that collide with the $\mu$MET with sufficient energy \citep{draineApJ79}.
In addition, we present here a more detailed analysis of how the considered physical processes evolve during the $\mu$MET flight.}.
Tab. \ref{max_surv_speed_tab} summarizes the maximum values of $v_\mathrm{in}$ that permits the survival of $\mu$METs with different initial radii, incidence angles, and densities. It is evident that 1) smaller $\mu$METs can have greater values of entry speed than greater $\mu$METs; 2) less dense $\mu$METs can sustain greater entry speeds than denser $\mu$METs and 3) $\mu$METs with large entry angle (i.e. far from the vertical) can survive higher entry speed than $\mu$METs with small entry angle (i.e. close to the vertical).

\begin{table}[h!]
\centering
\caption{Maximum values of entry velocity for which $\mu$METs are able to survive and reach the Earth surface.}
\begin{tabular}{c c c c c}
\hline
\mbox{\boldmath $r_\mathrm{in}$} & \multicolumn{4}{c}{\mbox{\textbf{max} \boldmath $v_\mathrm{in}$} \textbf{(km/s)}} \\ 
(\mbox{\boldmath $\mu$}\textbf{m)} & \multicolumn{3}{c}{$\rho_\mathrm{\mu MET} = 3$ g/cm$^3$}  & $\rho_\mathrm{\mu MET} = 1$ g/cm$^3$ \\ 
   & $z_\mathrm{in}$ = 0$^{\circ}$ & $z_\mathrm{in}$ = 45$^{\circ}$ & $z_\mathrm{in}$ = 70$^{\circ}$  & $z_\mathrm{in}$ = 45$^{\circ}$ \\
\hline
25  & 16   & 17   & 22 & 22 \\
50  & 15   & 16   & 21 & 21 \\
100 & 13   & 14   & 20 & 20 \\
200 & 12   & 13   & 18 & 18 \\
300 & 11.2 & 12   & 16 & 16 \\
400 &  -   & 12   & 15 & 15 \\
500 &  -   & 11.2 & 14 & 14 \\
\hline
\end{tabular}
\label{max_surv_speed_tab}
\end{table}

\begin{figure}[t]
\centering
  \resizebox{8.5cm}{!}{\includegraphics{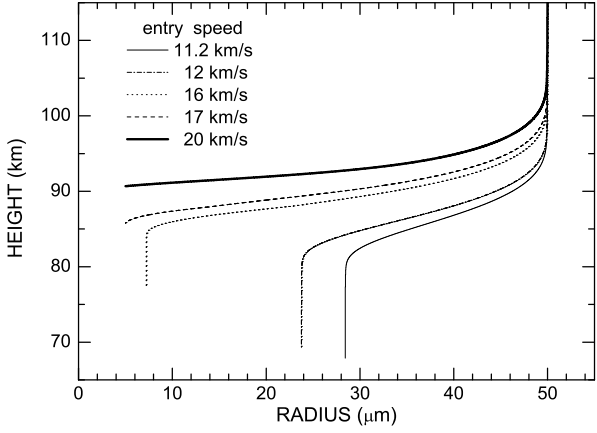}}	
  \caption{Evolution of $\mu$MET radius with height assuming $r_\mathrm{in}$ = 50 $\mu$m, $z_\mathrm{in}$ = 45$^{\circ}$ and $\rho_\mathrm{\mu MET}$ = 3 g/cm$^3$.  }
  \label{radius_vs_height}
\end{figure}

Figure \ref{radius_vs_height} (all figures refer to $\mu$METs with $\rho_\mathrm{\mu MET} = 3$ g/cm$^3$) shows the radius evolution of a 50 $\mu$m  $\mu$MET for  different values of $v_\mathrm{in}$. In the case of $z_\mathrm{in}$ = 45$^{\circ}$, for $v_\mathrm{in} = 11.2$ km/s the final $\mu$MET size is 28 $\mu$m, i.e. the $\mu$MET lost $\sim$82\% of its initial mass, while for $v_\mathrm{in} = 16$ km/s the final $\mu$MET radius is 7 $\mu$m (the mass lost is 99\%).  
Another important feature evident in Fig. \ref{radius_vs_height} is that $\mu$METs experience most of their alteration within a limited height interval ($\sim$15 km for the cases displayed in Fig. \ref{radius_vs_height}), corresponding to a duration of  only few seconds.

\begin{figure}[h]
\centering
  \resizebox{8.5cm}{!}{\includegraphics{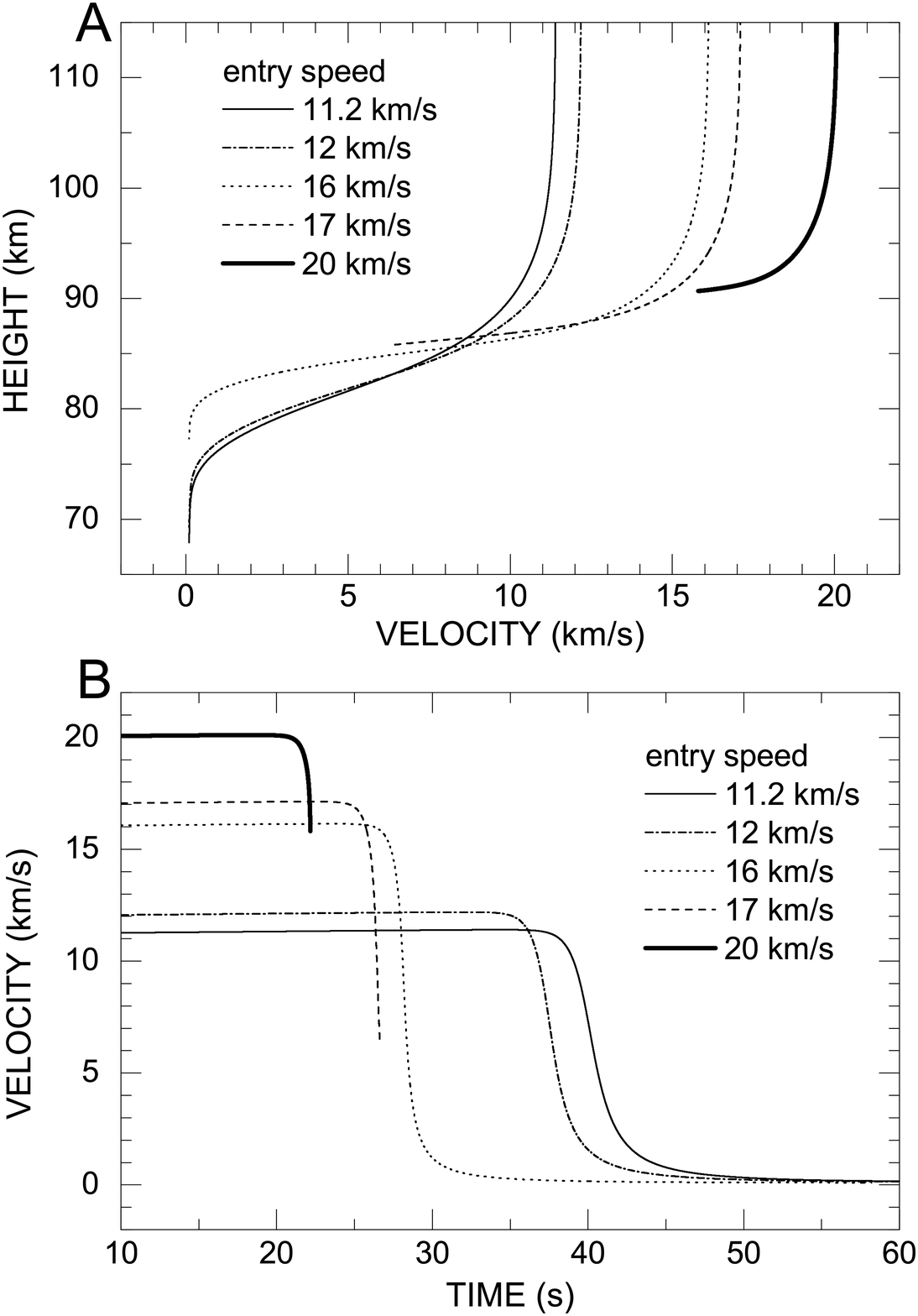}}	
  \caption{Evolution of $\mu$MET velocity with height (A) and time (B) assuming $r_\mathrm{in}$ = 50 $\mu$m, $z_\mathrm{in}$ = 45$^{\circ}$ and $\rho_\mathrm{\mu MET}$ = 3 g/cm$^3$.  }
  \label{speed_evol}
\end{figure}

The same is true for the time evolution of $\mu$MET's velocity. After a first phase in which the velocity remains almost constant, $\mu$METs are suddenly decelerated when they encounter sufficiently dense atmospheric layers (Fig. \ref{speed_evol}).
The height at which $\mu$METs start to be decelerated is higher for larger entry radii and higher initial velocities but the time interval in which $\mu$METs change from their initial to their final velocity is generally short.
A $\mu$MET with $r_\mathrm{in} = 50$ $\mu$m, $v_\mathrm{in} = 11.2$ km/s, $z_\mathrm{in}$ = 45$^{\circ}$ and $\rho_\mathrm{\mu MET} = 3$ g/cm$^3$ is slowed down to 1 km/s in $\sim$4 s, and then reaches its final velocity of $\sim$100 m/s within $\sim$30 s.  In contrast, for $r_\mathrm{in} = 300$ $\mu$m, $v_\mathrm{in} = 12$ km/s, $z_\mathrm{in}$ = 45$^{\circ}$ and $\rho_\mathrm{\mu MET} = 3$ g/cm$^3$, a $\mu$MET is slowed down to 1 km/s in only $\sim$2.5 s.  
Those $\mu$METs with grazing trajectories or lower density experience lesser variation. 
For a very oblique entry, they decelerate more gradually. 
In particular, the $\mu$MET's velocity (and therefore the rate of collisions) is reduced before the beginning of intensive ablation. Hence the maximum temperature is lower and the final radius is greater than for more nearly vertical entry.
Similarly, $\mu$METs with low density values (1 g/m$^3$) are decelerated at higher altitude where the atmospheric density is low.

\begin{figure}[!ht]
\centering
  \resizebox{8.5cm}{!}{\includegraphics{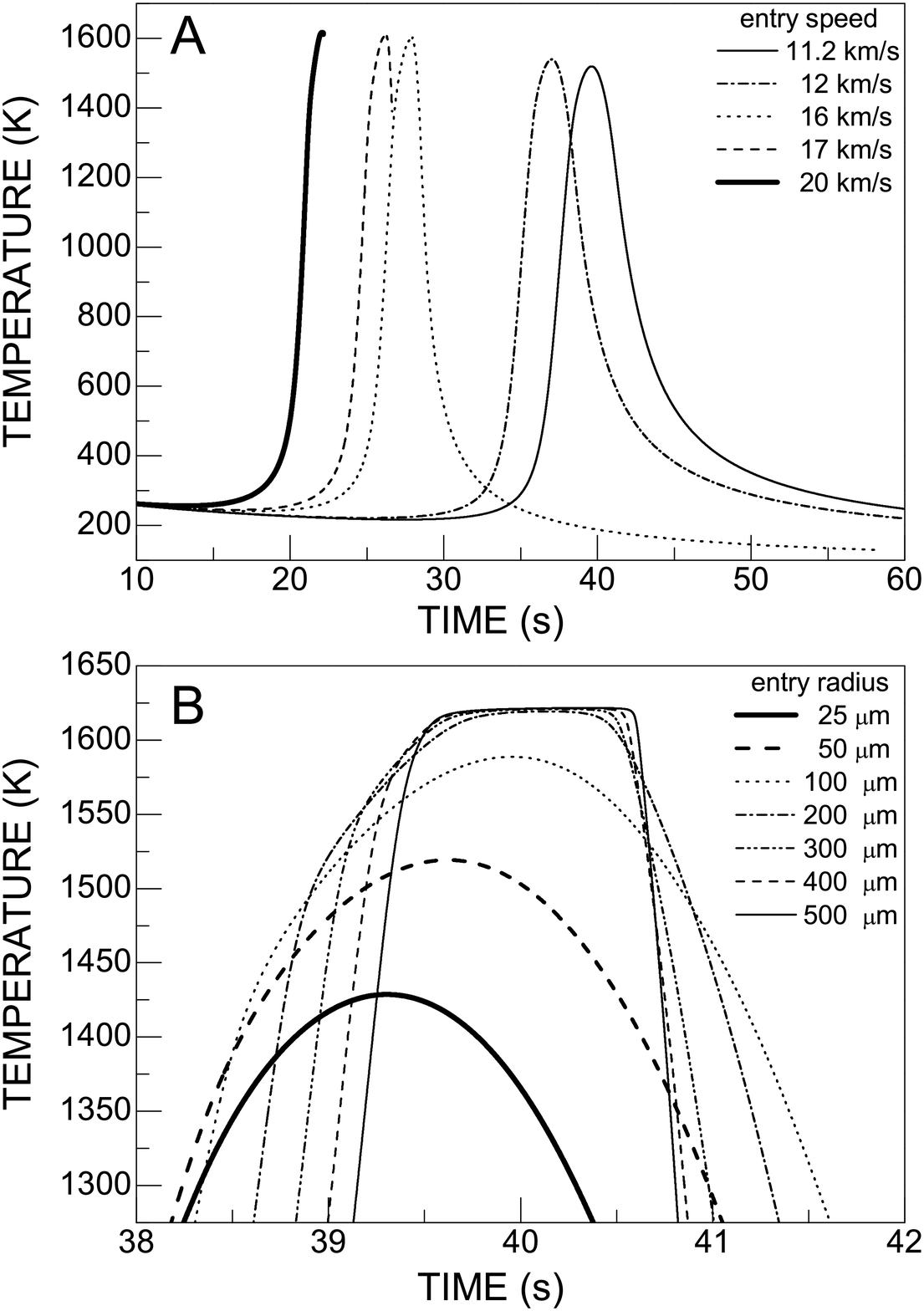}}
  \caption{Typical time variation of the $\mu$MET temperature.   A: results for $r_\mathrm{in} = 50$ $\mu$m, $z_\mathrm{in}$ = 45$^{\circ}$ and $\rho_\mathrm{\mu MET} = 3$ g/cm$^3$.  B: results for $v_\mathrm{in} = 11.2$ km/s, $z_\mathrm{in}$ = 45$^{\circ}$ and $\rho_\mathrm{\mu MET} = 3$ g/cm$^3$.   The truncated dashed and solid curves correspond to $\mu$METs that do not survive atmospheric passage. See text for discussion.}
  \label{temp_vs_time}
\end{figure}

Typical thermal histories are shown in Fig. \ref{temp_vs_time}A.  The almost impulsive temperature rise is followed by cooling with the whole process lasting only a few seconds.  The initial slight temperature decrease is due to efficient radiative emission and to the scarcity of collisions in the highest strata. $\mu$MET temperature would not decrease in the first phases of atmospheric entry if we included solar and atmospheric radiative heating  but, as previously discussed,  this contribution has negligible effects on $\mu$MET peak temperatures and final sizes (section \ref{sec:model_equations}).  At $h$ $\sim$100 -- 120 km, the $\mu$MET temperature rapidly rises because of  collisions, then   
after the maximum, $T_\mathrm{\mu MET}$ decreases very rapidly because the collisional heating is reduced as $\mu$METs slow down and shrink.  Taking the full width at half the maximum (FWHM) temperature to compare the duration of the heating pulse, $\mu$METs in Fig. \ref{temp_vs_time}A suffer heating pulses of 5.78 s (for $v_\mathrm{in} = 11.2$ km/s), 5.25 s (for $v_\mathrm{in} = 12$ km/s) and 3.15 s (for $v_\mathrm{in} = 16$ km/s).
Fig. \ref{temp_vs_time}A shows that peak temperatures are reached earlier for higher entry velocities but the altitude of peak temperature is about the same: 86 km for $v_\mathrm{in} = 11.2$ km/s, 88 km for $v_\mathrm{in} = 16$ km/s.  Fig. \ref{temp_vs_time}A and B show that higher entry speeds as well as larger initial radii imply higher peak temperature, but the differences are small.  Comparing Fig. \ref{temp_vs_time}A with Fig. \ref{radius_vs_height} shows that the  final radius is most sensitive to differences in entry speed.  This is due to the melting process that both limits the maximum temperature values (Fig. \ref{temp_vs_time}B) and generates significant mass loss.

\begin{figure}[]
\centering
  \resizebox{8.5cm}{!}{\includegraphics{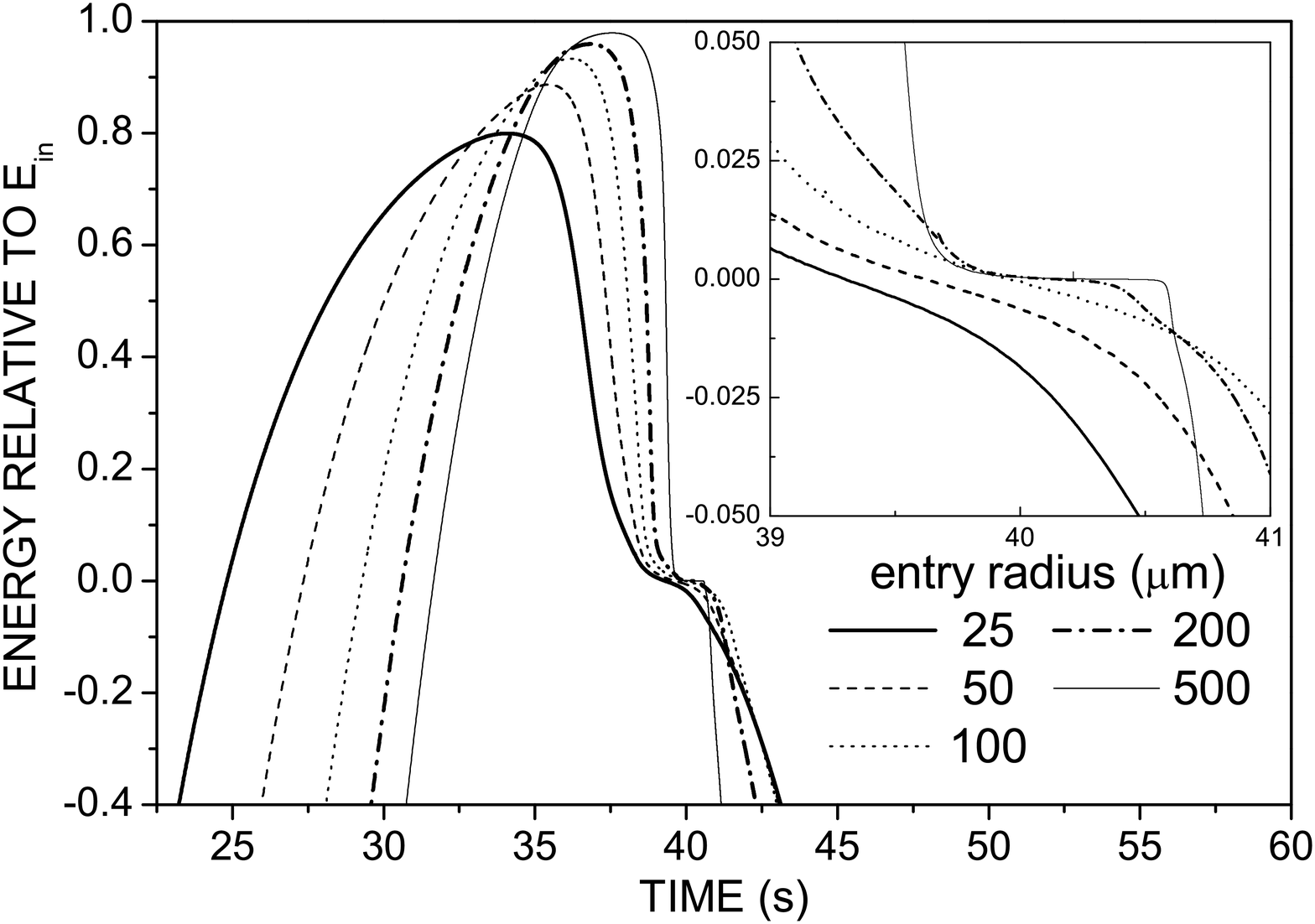}}
  \caption{Fraction of the absorbed energy used to change the $\mu$MET temperature. These curves are proportional to $dT_\mathrm{\mu MET}/dt$ so negative values mean that temperature is decreasing, positive values that it is increasing. The insert shows the transition from positive to negative values at around 40 s, coinciding with peak temperature and the start of $\mu$MET cooling by intense ablation.}
  \label{energy_vs_time}
\end{figure}

Fig. \ref{temp_vs_time}B shows temperature profiles for different values of the initial radius (with $v_\mathrm{in} = 11.2$ km/s).
Again, higher peak temperatures are reached for larger $r_\mathrm{in}$. 
Fig. \ref{temp_vs_time}B shows the clear difference between two classes of profiles. For $r_\mathrm{in} \leq 100$ $\mu$m the temperature curves have a regular, rounded shape.
In contrast, for $r_\mathrm{in}$ between 200 and 500 $\mu$m the top of the curves is due to melting of $\mu$MET material.  This is a general characteristic of our results.  It is not linked only to different entry radii, but it appears for several combinations of entry conditions.
These two classes of temperature profiles imply different final states for $\mu$METs.  
When the peak temperature is below the melting point and the temperature profile does not flatten, then the $\mu$MET survives after experiencing minor melting.  
These could correspond to the recovered unmelted and partially melted micrometeorites (see e.g. \citealt{gengeMAPS08}).
The other class of temperature profiles, with flatten peak near the melting point, correspond to $\mu$METs for which most of the mass is melted.
The difference between these two classes can be seen also analyzing the history of the energy producing the temperature variation, i.e. the term $cM_\mathrm{\mu MET}(dT_\mathrm{\mu MET}/dt)$ included in the energy balance. 
This term is displayed in Fig. \ref{energy_vs_time},  normalized to the incoming energy (equation \ref{eq_incoming_energy}), for the same cases as Fig. \ref{temp_vs_time}.  
These curves cross zero around 40 s, i.e. when the $\mu$MET temperatures reach their peak value and then start to decrease. As shown in the insert in Fig. \ref{energy_vs_time}, when $\mu$METs reach the melting temperature there is an interval during which no energy is used to alter their temperature. In contrast, curves corresponding to unmelted or partially melted $\mu$METs vary smoothly.

\begin{figure}[]
\centering
  \resizebox{8.5cm}{!}{\includegraphics{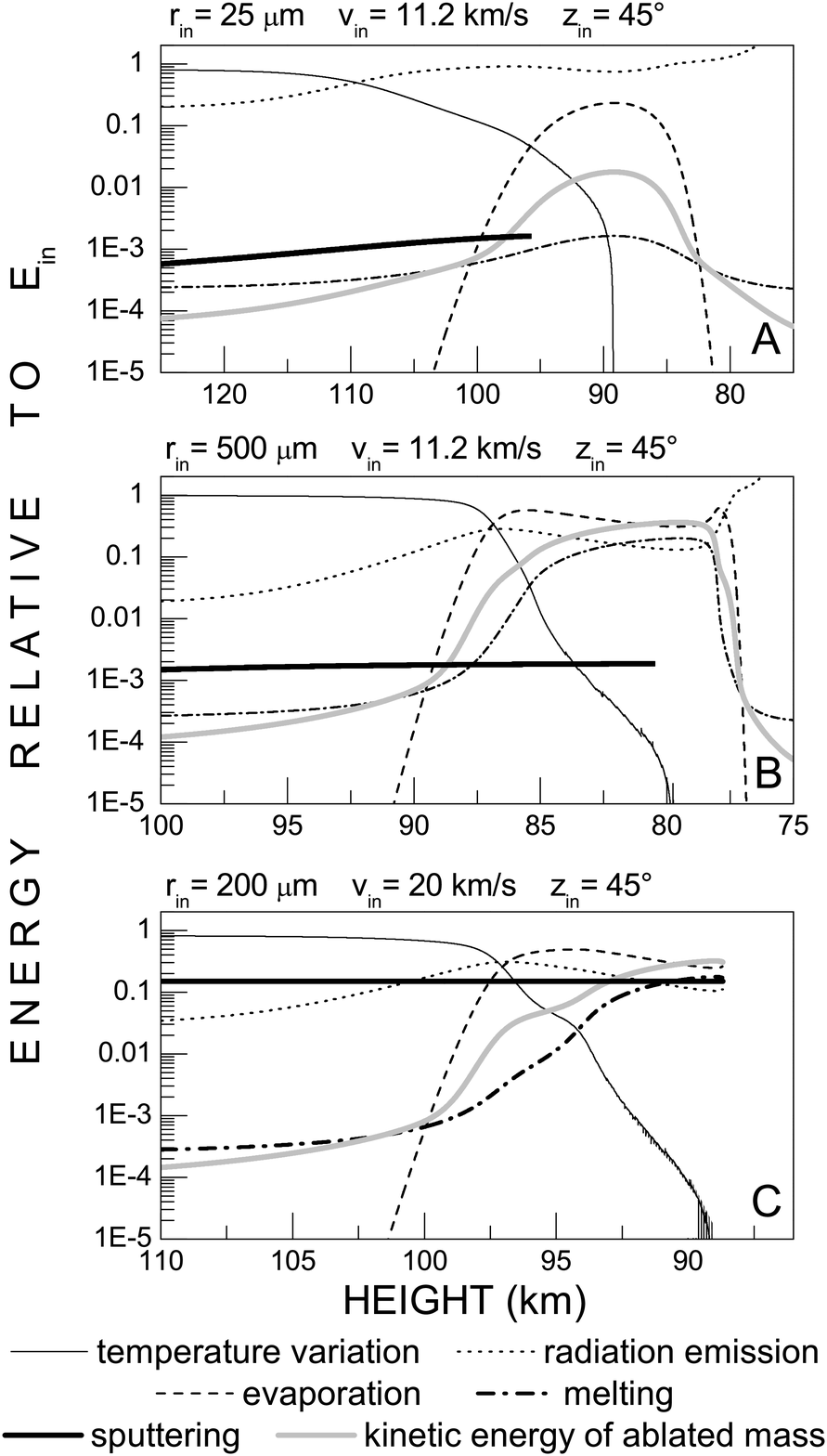}}
\caption{Variation of the energy going into different processes relative to the absorbed energy. A: unmelted or partially melted $\mu$MET. B: almost completely melted $\mu$MET.  C: completely destroyed $\mu$MET. See text for discussion. Noise in the temperature variation lines is due to the numerical integration and occurs when the calculation is performed on very small step sizes (of the order of 10$^{-5}$ s, used when $\mu$MET size, velocity and temperature vary significantly).}
  \label{en_vs_time_triple}
\end{figure}

Figure \ref{en_vs_time_triple} shows the distribution of the energy that $\mu$METs gain from collisions in the different processes considered in the energy balance.  In the first part of the trajectory most of the energy goes into raising the $\mu$MET's temperature (black, thin line).  At high temperature, $\gtrsim$1000 K, ablative processes (evaporation and melting) and the kinetic energy of the ablated mass become important.
However,  the relative importance of the different processes change with different entry conditions.
For the smallest $\mu$METs (Fig. \ref{en_vs_time_triple}A) the radiative losses dominate. This happens in two phases. At the very beginning, not shown in Fig. \ref{en_vs_time_triple}, when $\mu$METs suffer relatively few collisions,  their temperature ($T_\mathrm{in}$ = 330 K) is sufficient to produce significant emission. Then when the $\mu$MET speed and radius are greatly reduced, so is the energy acquired by collisions, the $\mu$MET temperature is near its peak values, and the radiative loss dominates.

\begin{figure}[t]
\centering 
  \resizebox{8.5cm}{!}{\includegraphics{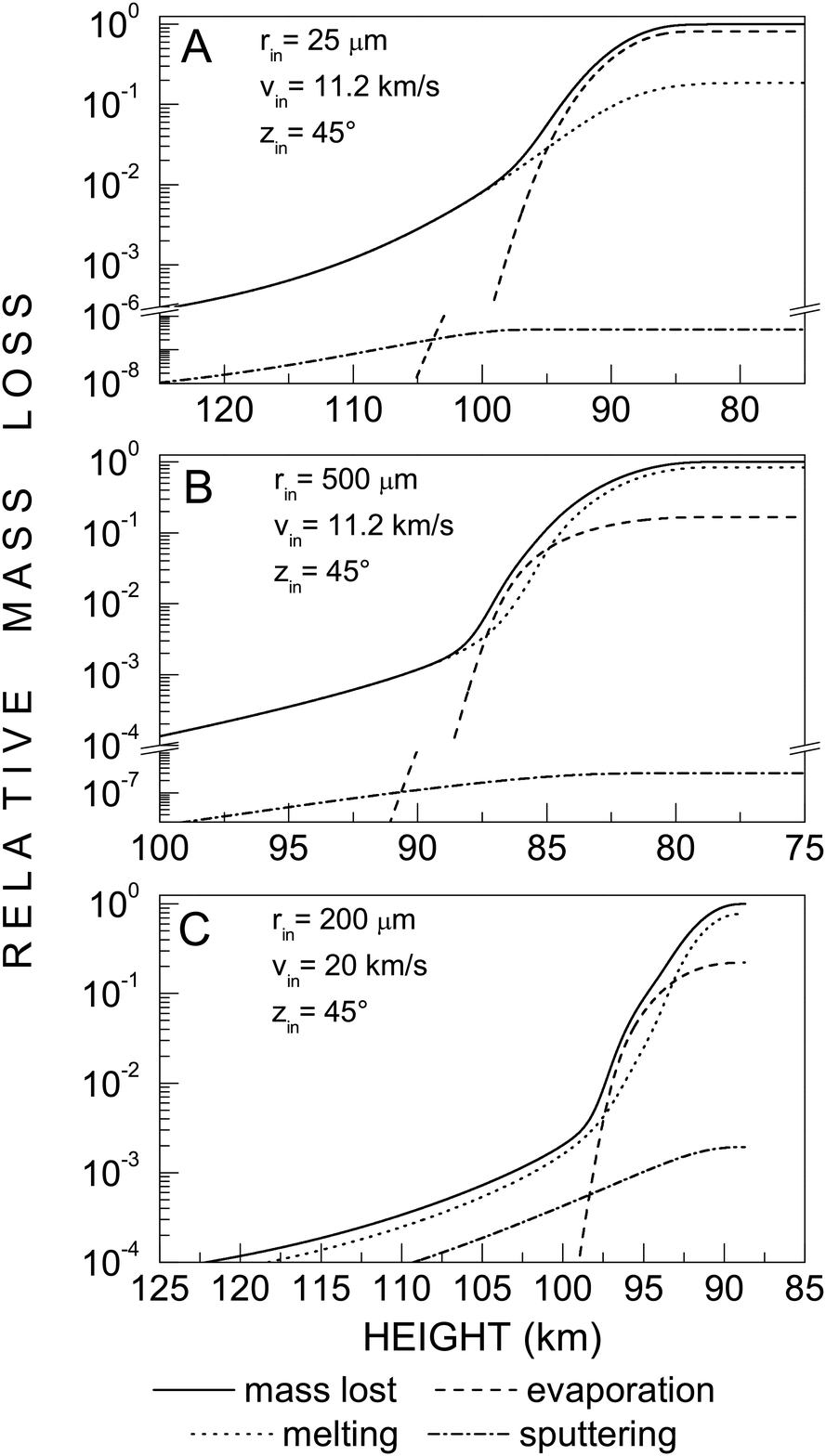}}
  \caption{Relative contribution to the mass loss of processes. The solid line (labelled ``mass lost'') shows the fraction of the total mass lost. A: evaporation dominates the mass loss, the final mass of this $\mu$MET is 47\% of the initial mass. B: melting dominates the mass loss, the final mass is $<$1\% of the initial mass. C: a $\mu$MET that is completely destroyed by melting.}
  \label{mass_loss}
\end{figure}

Fig. \ref{en_vs_time_triple}A shows the typical case for a partially melted or unmelted $\mu$MET. 
In the case shown, melting accounts for only about 0.1\% of the incoming energy, while up to about 25\% goes into evaporation. 
Sputtering is negligible, reaching about the same level as melting and terminating abruptly when $\mu$METs have slowed enough that the incident kinetic energy of atmospheric particles is below the sputtering threshold. 
For the case shown in Fig. \ref{en_vs_time_triple}B ablation processes (evaporation and melting) are more important than in the previous case.  
There is a time interval during which both evaporation and melting consume more energy than radiative losses.  Consequently, more mass is lost (see Fig. \ref{mass_loss}) and the kinetic energy removed by the ablated mass is important for the energy balance.  
Sputtering is still unimportant because of the low initial velocity ($v_\mathrm{in}$ = 11.2 km/s in Fig. \ref{en_vs_time_triple}B).  Instead, for $v_\mathrm{in}$ = 14 km/s (with $r_\mathrm{in}$ = 500 $\mu$m, $z_\mathrm{in}$ = 70$^{\circ}$ and $\rho_\mathrm{\mu MET}$ = 3 g/cm$^3$), we find an increase of about two order of magnitude, with the fraction of incident energy going into sputtering reaching $\sim$10\%.  
This is also found for $\mu$METs that do not survive atmospheric passage, as shown in Fig. \ref{en_vs_time_triple}C; sputtering accounts for more than 10\% in the energy balance, and melting and evaporation are more important than radiative losses.

The analysis of the mass balance (Fig. \ref{mass_loss}) confirms what we find from the energy balance. In Fig. \ref{mass_loss}A (unmelted or partially melted $\mu$MET) melting accounts for $\sim$20\% of the total mass lost. For these $\mu$METs, mass loss is mainly from evaporation; sputtering is negligible. 
The situation is opposite for the case shown in Fig. \ref{mass_loss}B: melting is the main cause of  mass loss, while evaporation driven mass loss is much lower ($\sim$16\% for the case of Fig. \ref{mass_loss}B).
In summary, sputtering is significant for greater entry velocities but, in our models,  {\it it is never the principal mechanism for mass loss}. In the case of Fig. \ref{mass_loss}C (relative to a $\mu$MET completely destroyed) sputtering is responsible for $\sim$0.1\% of the total mass lost.
Even in the most extreme case, $v_\mathrm{in}$ = 70 km/s and $r_\mathrm{in}$ = 500 $\mu$m, it accounts  for a modest 1\% of the mass loss.


\section[Discussion]{Discussion}
\label{sec:discussion}

We have assumed constant values for both the drag  $\Gamma = 1$ (equation \ref{eq_motion}) and the heat transfer  $\Lambda$ (equation \ref{eq_incoming_energy}) coefficients, a choice that is valid in a free molecular flow regime [see e.g. \citet{sorasioPSS01} and \citet{popovaASR07}] at high ${\cal M}_{MM}$.  
These coefficients are, however, dependent on the $\mu$MET temperature and the $\mu$MET Mach number  \citep{hoodICARUS91,meloshLPSC08}.  Thus, to test the sensitivity of our model to these parameters we performed simulations with different values of $\Gamma$ and $\Lambda$ for $r_\mathrm{in}$ = 100 $\mu$m, $v_\mathrm{in}$ = 12 km/s, $z_\mathrm{in}$ = 45$^{\circ}$ and $\rho_\mathrm{\mu MET}$ = 3 g/cm$^3$.  
Increasing $\Gamma$ means the $\mu$METs are decelerated more effectively, so they are less heated by collisions, reach lower peak temperatures (peak temperature for $\Gamma$ = 2 is 55 K lower than for $\Gamma$ = 1) and have larger final radii ($r_\mathrm{fin}$ = 67 $\mu$m for $\Gamma$ = 2, while $r_\mathrm{fin}$ = 30 $\mu$m for $\Gamma$ = 1).  
In contrast, for $\Gamma <$ 0.6 our test $\mu$MET is destroyed; for $\Gamma =$ 0.6 its peak temperature is 15 K higher than in the case $\Gamma$ = 1 and the final radius is 6 $\mu$m.  We find that $\mu$MET peak temperatures and final radii are more dependent on $\Lambda$.  Decreasing $\Lambda$ from 1 to 0.1 yields lower peak temperatures, decreasing from 1604 K to 1255 K.  
The final radius is 99 $\mu$m for $\Lambda$ = 0.1, i.e. $\mu$MET size is unaffected by the interaction with the atmosphere as also found by \citet{campbell-brownAA04}.  However, such low values of $\Lambda$ are unlikely for the cases considered here. To estimate at what $\Lambda$   the $\mu$METs experience the most intense alteration at peak temperatures, we use the expression of \citet{hoodICARUS91}:
$$
\Lambda = 1 - \frac{1}{2(v_\mathrm{\mu MET}/v_\mathrm{th})^2}\left(\frac{\gamma+1}{\gamma-1}\right)T_\mathrm{\mu MET}/T_\mathrm{air}
$$
where $\gamma$=7/5 is the ratio of the specific heats of the terrestrial atmospheric gases and $T_\mathrm{air}$ is the atmospheric temperature. Since $v_\mathrm{th}$ is computed as $v_\mathrm{th}= \sqrt{3k_\mathrm{B}T_\mathrm{air}/m_\mathrm{avg}}$ ($k_\mathrm{B}$ = Boltzmann constant and $m_\mathrm{avg}$ = mean mass of atmospheric particles) the value of $\Lambda$ is determined by the ratio $T_\mathrm{\mu MET}/v_\mathrm{\mu MET}^2$. 
In our simulations the maximum  $T_\mathrm{\mu MET}$ is always reached before $v_\mathrm{\mu MET}$ is significantly reduced, and when $v_\mathrm{\mu MET}$ decreases, so does $T_\mathrm{\mu MET}$.  
Thus, $\Lambda$ remains close to 1.   This is illustrated in Table 2.

\begin{table}[h!]\footnotesize
\centering
{\renewcommand{\arraystretch}{1.5}
\renewcommand{\tabcolsep}{0.2cm}
\caption{Values of the heat transfer coefficient $\Lambda$ for three test case $\mu$METs with z$_\mathrm{in}$= 45$^{\circ}$ and $\rho_\mathrm{\mu MET}$= 3 g/cm$^3$. Simulations performed in the present-day atmosphere.}
\begin{tabular}{l l l l}
\hline
 Parameter & Case 1 & Case 2 & Case 3\\
\hline
$r_\mathrm{in}$ ($\mu$m) & 50  & 100  &  500 \\
$v_\mathrm{in}$ (km/s) & 16 & 12 & 11.2 \\
$T_\mathrm{\mu MET}^\mathrm{max}$ (K) & 1606 & 1603 & 1622  \\
$v_\mathrm{\mu MET}$ (km/s) at $T_\mathrm{\mu MET}^\mathrm{max}$ & 12.45 & 9.65 & 10.0  \\
$h$ (km) at $T_\mathrm{\mu MET}^\mathrm{max}$ & 88.0 & 84.1 & 79.5  \\
$T_\mathrm{air}(h)$ (K) & 176.2 & 181.4 & 191.6 \\
$v_\mathrm{th}(h)$ (m/s) & 392 & 398 & 409 \\
$\Lambda$ & 0.973 & 0.955 & 0.958 \\
\hline
\end{tabular}} 
\label{test_lambda}
\end{table}

The coefficient $\eta$, the fraction of the absorbed energy going into sputtering, is also important but choosing its value is very difficult since previous theoretical (e.g. \citealt{Sigmund81}) and experimental (e.g. \citealt{Nakles04}) studies have treated sputtering as an isolated process.
For instance, the mean mass of $\mu$MET atoms and their binding energy are necessary parameters for specifying the sputtering yield $Y$, and its  contribution to the energy balance.  
We note that a simplified treatment of the structure and composition of the $\mu$METs, as in this work, only approximates the sputtering contribution and this process is probably more important for less compact $\mu$METs than those treated here.
Any treatment of the sputtering yield and heating hinges on several critical points: the evaluation of the different processes that cause the removal of a particle from a solid body \citep{mayMNRAS00}, the sputtering yield dependence on the projectile incidence angle \citep{juracApJ98}, which particles are ejected from the target (atoms? molecules? both? See \citealt{tielensApJ94}) and the speed at which they leave the target \citep{popovaASR07}.
\citet{coulsonMNRAS02} assumed a Maxwellian distribution for the speed of sputtered particles.  This produces a momentum transfer to $\mu$METs about 11 times greater than the case of collisions alone, with consequent significant decreases of the $\mu$MET speed, peak temperature and ablation.  In our model, we do not include any impulsive reaction sustained by $\mu$METs from removed fragments since we assume isotropic ablative mass loss, although we do account for their kinetic energy in the energy balance.  

Finally the atmospheric component number density also affects sputtering.  
We find that sputtering-induced mass loss is always a small fraction ($\leq$0.01) of the total amount.
This contrasts with the results of \citet{rogersPSS05}, who find that sputtering accounts for up to about 40\% of the mass loss for $\mu$METs with $\rho_\mathrm{\mu MET} = 3$ g/cm$^3$ and up to half for $\mu$METs with $\rho_\mathrm{\mu MET} = 1$ g/cm$^3$.  
We suggest that the difference between our results and those obtained by \citet{rogersPSS05} is due to their assumption that all the energy transferred to $\mu$METs by collisions with atmospheric particles produces sputtering, rather than only a fraction of it (i.e. they assume $\eta$ = 1) and to their overestimate of the number densities of the atmospheric components that are systematically two order of magnitude greater than those we obtained from the same atmospheric model (the MSISE-90 model).

Our model assumes that $\mu$METs are homogeneous and that all have the same composition. Recovered micrometeorites are, however,  complex aggregates of different phases (e.g. silicates, sulfides, carbonaceous material) and show a wide range of compositions.
More realistic $\mu$METs could be modeled assuming different proportions of the constituent phases.
This would require modeling $\mu$METs of different densities and with different melting temperatures.
Our results show that a lower $\mu$MET density favors their survival (see Table \ref{max_surv_speed_tab}) because they are decelerated at higher altitude, in less dense atmospheric layers.
Qualitatively, we expect that lower melting temperatures correspond to those $\mu$METs that are mainly composed of more fragile phases, i.e. phases easily removed by the atmospheric frictional heating, as carbonaceous material and that higher melting temperatures correspond to those mainly composed of refractory phases, such as silicates, that are less affected by the atmospheric frictional heating.
\citet{vondrakACP2008} modeled the differential ablation of the $\mu$MET constituents, but without considering heterogeneities and retaining the hypothesis of uniform $\mu$MET temperature.  We too assumed that $\mu$METs have a uniform temperature at each instant of their atmospheric flight, as in previous models \citep{loveICARUS91,campbell-brownAA04}.  
The problem of internal temperature gradients was treated by \citet{szydlikLPSC1992},\citet{flynnLPSC95},\citet{szydlikMetSoc1997},\citet{gengeLPSC2000}. However, in all these studies, the $\mu$MET radius is kept constant as are the thermal conductivity and heat capacity (i.e. they assume homogeneous composition). A model for the simulation of the atmospheric passage of heterogeneous, multiphase $\mu$METs and of their internal temperature distribution would be a major advance.


\section[Conclusions]{Conclusions}
\label{sec:conclusion}

To summarize, we list here our major conclusions.  For any initial radius, a higher entry velocity and a more nearly normal incidence yield more severe heating for $\mu$METs, and therefore greater destruction rate. For fixed entry velocity and angle, the larger the radius, the greater the amount of mass lost.
For all the $\mu$METs simulated in this work radiation emission is the most important process at greatest altitude, before $\mu$METs encounter atmospheric layers sufficiently dense to activate ablation processes (evaporation and melting).
Melting is the most effective process that limits the increase of the $\mu$MET temperature and produces significant mass loss.  In all cases, the contribution of sputtering is found to be negligible in both the energy and the mass balance but could be more important for more fragile $\mu$METs than those simulated here.  
For the smallest $\mu$METs simulated in this work ($r_\mathrm{in} = 25 \mu$m), radiation is the dominant cooling process during all their atmospheric flight, allowing the survival of these small $\mu$METs.  
Ablative processes are more important for larger $\mu$METs, i.e. these  lose a larger fraction of their mass than small $\mu$METs and only low entry velocities and grazing entry angles insure survival during atmospheric passage.

Our model shows that, among $\mu$METs that are not completely destroyed, two types can be distinguished based on their temperature evolution. The first group is those $\mu$METs that show smooth temperature curves, i.e. that are not completely melted during atmospheric flight and therefore reach the surface as unmelted or partially melted micrometeorites (most of the mass lost by these $\mu$METs is due to evaporation).  
The second comprises $\mu$METs with flattened temperature curves near the maximum temperature.
As we have shown, these differences are due to the importance of the melting process in the energy and the mass balance equations.

We view this  model as a first step in a broader study of $\mu$METs-atmosphere interactions. The necessary refinements require a broadly  multidisciplinary approach that combines the physics with laboratory studies of the material properties from chemistry and mineralogy, for the alteration sustained by  $\mu$METs in flight, with detailed numerical hydrodynamic simulations of the composite materials typical of real samples. 
Finally, the analysis of the biochemical aspects, i.e. related to the organic molecules present in $\mu$METs, could improve the understanding of what role $\mu$METs played in the emergence of life on Earth (and, by extension, extrasolar planets).  For example, although its is beyond the scope of the present paper, this type of modeling can be applied to the case of the interaction of $\mu$METs with the primordial terrestrial atmosphere to assess the role of $\mu$METs as conveyors of pre-biotic molecules \citep{brianiPhD2010}.

\section*{Acknowledgments}
\label{sec:aknowl}

We thank M. D'Orazio and L. Folco for valuable discussions and F. Savini for assistance. We thank F. Rietmeijer for a careful review of a first version of the manuscript. This work was supported in part by MIUR.

\bibliographystyle{aa}
\bibliography{bibliography_SAMBA_formatted}

\end{document}